\documentclass[12pt]{iopart}

\usepackage{iopams}
\usepackage{xcolor}
\usepackage[pdftex]{graphicx}
\usepackage{amsthm}
\begin{document}

\title[Detecting entanglement of unknown CV states with random measurements]{Detecting entanglement of unknown continuous variable states with random measurements}

\author{Tatiana Mihaescu (1,2), Hermann Kampermann (1), Giulio Gianfelici (1), Aurelian Isar (2,3), Dagmar Bru{\ss} (1)}

\address{(1) Heinrich-Heine-Universit\"at D\"usseldorf, Institut f\"ur Theoretische Physik III, D-40225 D\"usseldorf, Germany

(2) Department of Theoretical Physics, National Institute of Physics and Nuclear Engineering, RO-077125 Bucharest-Magurele, Romania

(3) Faculty of Physics, University of Bucharest, RO-077125 Bucharest-Magurele, Romania}
\ead{mihaescu.tatiana@theory.nipne.ro}
\vspace{10pt}

\begin{abstract}
We explore the possibility of entanglement detection in continuous variable systems by entanglement witnesses based on covariance matrices, constructible from random homodyne measurements. We propose new linear constraints characterizing the entanglement witnesses based on second moments, and use them in a semidefinite program providing the optimal entanglement test for given random measurements. We test the method on the class of squeezed vacuum states and study the efficiency of entanglement detection in general unknown covariance matrices.
\end{abstract}

\section{Introduction}

The most valuable characteristics of quantum systems are quantum correlations such as entanglement, which represents a useful resource for applications unattainable in the framework of classical theory, such as quantum teleportation, quantum cryptography and dense coding \cite{hor, brus}. In the last two decades, the use of continuous variable (CV) systems became a very powerful approach to quantum information processing, opening the way to various protocols and tasks, like quantum cryptography, quantum teleportation, quantum state discrimination \cite{braun,sera,weed}. CV quantum systems provide the quantum description of the propagating electromagnetic field, and therefore manifest particular relevance for quantum communication and quantum techniques like detection, imaging and sensing.

A significant problem in quantum information theory and any application is to efficiently reveal the properties of an unknown quantum state, in particular to certify the presence of entanglement in a given unknown state. Usual entanglement criteria for CV systems consist in certain operations on the second moments, or uncertainties, of quantum states, such as the positive partial transpose (PPT) criterion \cite{per, sim}. Therefore, first a full tomography is required for completely unknown states, in order to reconstruct the entire covariance matrix. However, this method may be a very resource-consuming and demanding experimental procedure, especially for quantum states with a high number of modes. In addition, the full information about the second moments of the state can be excessive for the characterisation of entanglement present in the state. Instead, one can choose to measure certain fixed combinations of second moments, giving rise to a specific test, which detects entanglement in some states and does not detect it in others \cite{duan, aoki, look}.

Entanglement witnesses (EWs) represent another commonly used entanglement test, being directly accessible in experiments through measurable observables \cite{toth}. A Hermitian observable $W$ is an entanglement witness if for all separable states $\rho_s$, $\Tr[W\rho_s]\geq 0$ holds, while for some entangled state $\rho$ we have that $\Tr[W\rho]<0$ \cite{toth}. For CV systems a special instance of entanglement witnesses can be defined, which embodies the entanglement criterion in terms of the variances of the canonical observables of the state \cite{and, hyll, sper, shch}. Typically, entanglement witnesses are employed when certain knowledge about the state is available.

Given an unknown quantum state, however, the complexity of the state and the absence of any information about it deprive us of a specific experimental strategy in tackling the problem of efficient entanglement detection. Therefore, the best strategy in this case would be to perform random measurements, serving as building blocks for the construction of an entanglement witness by means of a semidefinite optimization algorithm. This idea is inspired by an analogous method for the discrete-variable case, which was developed in \cite{jochn}.

The paper is organized as follows. In Sec. 2 we present the theoretical framework of CV states, mainly based on the second moments description of the state. In Sec. 3 we introduce the entanglement witnesses based on the covariance matrix (CM) of the state, as presented in Refs. \cite{and, hyll}, and propose a set of stronger linear semidefinite constraints in order to characterize the EWs.  Then, we simulate random homodyne measurements for two-mode CMs in Sec. 4 and formulate a semidefinite program (SDP) optimizing the witness constructible from given experimental data. We present the results of the efficiency of entanglement detection for random two-mode CMs and, in particular, for the class of squeezed vacuum states in Sec. 5, and illustrate an example of  bipartite bound EW.  A statistical analysis of our method is provided in Sec. 6, and a summary and conclusions are presented in Sec. 7.

\section{Continuous variable systems}

A CV system of $N$ canonical bosonic modes, like the quantized electromagnetic field with a Hamiltonian of a system of  $N$ harmonic oscillators (modes), is defined in a Hilbert space $ \mathcal H=\bigotimes_{k=1}^N \mathcal H_k$, each one with an infinite-dimensional space $\mathcal H_k=L^2(\mathbb R)$ and two canonical observables $\hat x_k$ and $\hat p_k$, with the corresponding phase space variables of position $x_k$ and momentum $p_k$ \cite{braun, sera, weed}. One can define a vector of quadrature  operators $\hat R^{\rm T}\equiv(\hat R_1,...,\hat R_{2N})=(\hat x_1,\hat p_1,...,\hat x_N,\hat p_N)$ satisfying the bosonic commutation relations
\begin{equation}\label{cmr}
  [\hat R_i,\hat R_j]=i\Omega_{ij} \hat I,\quad i,j=1,...,2N,
\end{equation}
where $\hat I$ is the identity matrix, and $\Omega_{ij}$ are the elements of the fundamental symplectic matrix (we assume $\hbar=1$)
\begin{equation}\label{sym}
\Omega_N=\bigoplus_{1}^N \left({\begin{array}{*{50}c} 0 & 1 \\ -1 & 0 \\ \end{array}}\right).
\end{equation}
The primary role in this study is played by the statistical moments of the quadrature operators, that characterize the state with density operator $\rho$ \cite{braun,weed} , up to the second order: the displacement vector, which is the real vector  $d$ of first order moments $d_i=\langle \hat R_i\rangle_\rho=\Tr[\hat R_i\rho]$, and the covariance matrix (CM), which is the real, symmetric matrix $\gamma$ whose entries are the second order moments in symmetrized form (the variances) of the quadrature operators, defined as \cite{sera, mukunda}:
\begin{equation}\label{ptr}
\gamma_{ij}=\langle\{\hat R_i-\langle\hat R_i\rangle, \hat R_j-\langle\hat R_j\rangle\}_+\rangle_\rho,
\end{equation}
where $\{ ,\}_+$ represents the anticommutator. The Robertson-Schr\"odinger uncertainty relation in terms of the CM reads
\begin{equation}\label{unc}
\gamma+i\Omega_N \geq 0,
\end{equation}
assuring that it is a CM of a physical quantum state. Gaussian states represent the class of CV states which are completely characterized by their first and second moments. The entanglement criteria discussed in this paper can also detect entanglement of non-Gaussian states.

A quantum state of a bipartite system is entangled if it cannot be prepared by means of operations acting locally on the subsystems. In the case of separable states correlations are attributed to possible classical communication between subsystems, and hence are of classical origin. This reasoning  carries over to CV systems, where a separability criterion can be defined in terms of CMs. If a CM $\gamma$ of a state of $N$ modes is fully separable, then there exist CMs $\gamma_i$, $i=1,\ldots, N$, corresponding to $N$ subsystems, such that \cite{ww}
\begin{equation}
\gamma\geq \bigoplus_{i=1}^N \gamma_i.
\end{equation}
Conversely, if this holds, then Gaussian states with CM $\gamma$ are separable. Therefore, if this criterion is violated, then the corresponding state is entangled, irrespective of whether it is Gaussian or not. If it is not violated, then a Gaussian state is separable, while a non-Gaussian state might be entangled.\\
In the following, we will refer to the situation of a $k-$partition of an $N-$mode system as the splitting or distribution of an $N-$mode system into $k$ subsystems, where every subsystem $j$ ($j=1,\ldots,k$) is composed of $N_j$ modes, such that $\sum_{j=1}^k N_j=N$.\\

\subsection{Symplectic transformations}

Unitary operators acting on the quantum state space are equivalent to symplectic transformations which preserve the commutation relations of canonical variables. The real symplectic group is defined by \cite{dutta}:
 \begin{equation}
Sp(2N,\mathbb R)=\{S\in \mathcal{M}(2N,\mathbb R): S\Omega_N S^{\rm T}=\Omega_N \},
\end{equation}
where $S$ is a symplectic transformation acting in phase space as $\hat R \to \hat R'=S \hat R$, and $\mathcal{M}(2N, \mathbb R)$ denotes the set of $2N\times 2N$ real matrices. Symplectic transformations act by congruence on CMs: $\gamma'=S\gamma S^{\rm T}$.

Every symplectic transformation can be decomposed using the Euler decomposition, which represents the singular value decomposition for real symplectic matrices \cite{dutta}:
\begin{equation}\label{eu}
  S=K\Big[\bigoplus_{i=1}^{N} S(r_i) \Big] L,
\end{equation}
where $K$, $L$ are symplectic and orthogonal matrices, while
\begin{equation}\label{sqr}
  S(r_i)=\left({\begin{array}{*{20}c} e^{-r_{i}} & 0  \\
0 & e^{r_{i}} \end{array}}\right)
\end{equation}
are one-mode squeezing matrices (symplectic and nonorthogonal) with $r_i$ the squeezing parameter. The symplectic and orthogonal matrices form the maximal compact subgroup $K(N)$ within the noncompact group $Sp(2N,\mathbb R)$ \cite{dutta}. The group $K(N)$ is isomorphic to the group $U(N)$ of $N\times N$ complex unitary matrices:
  \begin{equation}\label{ku}
    K(N)=\{S(X,Y)|X-iY\in U(N)\},
  \end{equation}
  where the corresponding symplectic matrices are of the following form:
 \begin{equation}\label{sorth}
   S(X,Y)=\left({\begin{array}{*{20}c} X & Y  \\
-Y & X
\end{array}}\right) \in Sp(2N, \mathbb R).
\end{equation}
Such transformations describe multiport interferometers and are called passive canonical unitaries, which preserve the photon number \cite{weed}. The active canonical unitaries  correspond to nonorthogonal symplectic transformations, such as one-mode squeezers.

In the following  we will use the theorem by Williamson \cite{will}, according to which every matrix $M\in\mathcal{M}(2N,\mathbb R)$, $M\geq0$ can be brought to a diagonal form through symplectic transformations:
\begin{equation}
SMS^T={\rm diag}(s_1, s_1,...., s_N, s_N),
\end{equation}
where $s_1,....,s_N\geq 0$ are called symplectic eigenvalues of $M$. By
\begin{equation}\label{defstr}
{\rm str}[M]:=\sum_{i=1}^N s_i
\end{equation}
we will denote the symplectic trace of $M$.

\section{Entanglement witnesses for covariance matrices}

An entanglement witness based on CMs is characterised by a real symmetric matrix $Z\geq 0$ such that \cite{hyll}
\begin{eqnarray}\label{subeq}
&\Tr[Z \gamma_s]\geq 1, \qquad {\rm for~all~separable~CMs} \qquad \gamma_s, \nonumber\\
&\Tr[Z \gamma]<1, \qquad {\rm for~some~entangled~CM} \qquad \gamma.\label{12b}
\end{eqnarray}
The EWs based on second moments defined in Eqs. (\ref{12b})  represent hyperplanes in the space of CMs that separate some entangled states from the set of separable CMs. If there exists $Z$ which fulfills conditions (\ref{12b}), then the state with CM $\gamma$ is entangled, irrespective of whether it is Gaussian or not, while if this test does not detect entanglement in a given non-Gaussian state, then the result is inconclusive. The following Theorem fully characterises the EWs for multimode CV states defined in Eqs. (\ref{12b})  for different entanglement classes.

\textbf{Theorem.} (taken from \cite{and,hyll}) \textit{A covariance matrix $\gamma$ of a  $k-$partite system with $\sum_{j=1}^k N_j=N$ modes is entangled with respect to this partition iff there exists $Z$ such that
\begin{equation}\label{dett}
\Tr[Z \gamma]< 1,
\end{equation}
where $Z$ is a real, symmetric $2N\times 2N$ matrix satisfying
\begin{eqnarray}\label{wine}
& Z\geq 0, \nonumber\\
& \sum_{j=1}^k{\rm str}[Z_j]\geq \frac 1 2,
\end{eqnarray}
where $Z_j$ is the block matrix on the diagonal of $Z$ acting on the subsystem $j$. Matrices $Z$ are called EWs based on second moments.}

Due to the convexity of the set of separable CMs there always exists an EW $Z$ giving the result of Eq. (\ref{dett}) for an entangled CM $\gamma$. In Ref. \cite{hyll} the authors formulated the above theorem in a slightly different way: in addition to Eqs. (\ref{wine}) it is stated that such an EW has to satisfy also ${\rm str}[Z]<\frac 1 2$, instead of condition $\Tr[Z \gamma]< 1$. Note that there is no contradiction between the conditions (\ref{wine})  and ${\rm str}[Z]< \frac 1 2$ that an EW has to satisfy, since the relation ${\rm str} [Z]\leq\sum_{i=1}^N{\rm str}[Z_i]$ holds \cite{and}.\\
\indent Nevertheless, the two formulations of the Theorem above are equivalent. In order to show this we will use the results from Ref. \cite{and} where it is proven that $\Tr[Z\gamma]\geq 1$ for all separable CMs $\gamma$ if and only if $Z\geq 0$ and $\sum_{j=1}^k{\rm str}[Z_j]\geq \frac 1 2$. In addition,  it is shown that $\Tr[Z \gamma]\geq1$ for all CMs $\gamma$ if and only if $Z\geq 0$ and ${\rm str}[Z]\geq \frac 1 2$. As $Z\ngeq 0$ would contradict $\Tr[Z \gamma]\geq1$ for all separable CMs $\gamma$, it follows that if $\Tr[Z\gamma]<1$ for some CM $\gamma$ then ${\rm str}[Z]< \frac 1 2$. Conversely, if ${\rm str}[Z]< \frac 1 2$ then there exists some CM $\gamma$ such that $\Tr[Z\gamma]<1$.

The problem of finding an EW that most robustly detects entanglement in a given CM arises as a semidefinite optimization problem (SDP) (see Ref. \cite{hyll} where the authors provide also numerical routines performing this task). Here we consider the situation when no information about the state is available, and we aim at constructing the EWs from given random measurements. For this purpose, the description of EWs given in the Theorem above can serve as constraints in our optimization program.

However, the inequality (\ref{wine}) cannot be used in this form as a semidefinite constraint in an SDP, because its left hand-side cannot be regarded as a linear function since the symplectic eigenvalues of a matrix $M\geq 0$ are given by the eigenvalues of the matrix $M^{\frac 1 2}(i\Omega_{N}) M^{\frac 1 2}$ \cite{sera}. In the following, we propose a set of linear semidefinite constraints for EWs, which are stronger than conditions (\ref{wine}).

\textbf{Proposition.} \textit{For the entanglement witness $Z$ of a $k-$partite entangled $N-$mode covariance matrix  with $\sum_{j=1}^k N_j=N$, the inequalities (\ref{wine}) are satisfied if the following conditions are fulfilled:
\begin{eqnarray}\label{prop}
& Z\geq 0, \nonumber\\
& Z_j+i\frac{x_j}{N_j} \Omega_{N_j} \geq 0, \quad x_j\in \mathbb{R},\quad j=1,....,k-1,\\
& Z_k+i \frac{1}{N_k}\Big(\frac 1 2-\sum_{j=1}^{k-1}x_j\Big) \Omega_{N_k} \geq 0.\nonumber
\end{eqnarray}}

\begin{proof}
\begin{enumerate}
\item In the first part we prove the proposition for $k=N$. First, we prove this for $N=2$, i.e. for a two-mode witness with the following block form:
\begin{equation}\label{z2}
  Z=\left({\begin{array}{*{20}c}
      Z_1 & Z_c \\
      Z_c^{\rm T} & Z_2
    \end{array}}\right),
\end{equation}
where $Z_1$ and $Z_2$ are $2 \times 2$ matrices. Since $Z$ is a positive semidefinite matrix, also the principal submatrices $Z_1$ and $Z_2$ are positive semidefinite. Let us assume the following inequality:
\begin{equation}
Z_1+i x \Omega_1 \geq 0,\quad {\rm where} \quad  \Omega_1= \left({\begin{array}{*{20}c} 0 & 1  \\
-1 & 0
\end{array}}\right), \quad x\in \mathbb R.
\end{equation}
By symplectic transformation $S$ the positive matrix above can be brought to the Williamson normal form as follows \footnote{Any symplectic transformation preserves the symplectic eigenvalues, and since we know that ${\rm Tr}[M]\geq 2  \ {\rm str}[M]$ holds for any positive matrix $M$ \cite{bathia}, then we may say that symplectic transformations preserve also the positivity.}:
\begin{equation}\label{poz}
S(Z_1+i x  \Omega_1 )S^{T}=Z_1^w+i x  \Omega_1=\left({\begin{array}{*{20}c} z_1 & ix  \\
-ix & z_1
\end{array}}\right),
\end{equation}
where $Z_1^w={\rm diag}(z_1,z_1)$, with $z_1$ the positive symplectic eigenvalue of $Z_1$. The eigenvalues $\alpha$ of matrix (\ref{poz}) are determined from the equation:
\begin{equation}
(z_1-\alpha)^2-x^2 =(z_1-\alpha-x )(z_1-\alpha+x )=0,
\end{equation}
and hence
\begin{equation}
z_1\pm x=\alpha\geq 0.
\end{equation}
Thus, the symplectic eigenvalue $z_1$ fulfills the inequality $z_1\geq\pm x$, or $z_1\geq |x|$. A similar inequality can be formulated for the block matrix $Z_2$:
\begin{equation}
Z_2+i(\frac 1 2-x) \Omega_1 \geq 0,
\end{equation}
from which we obtain the following condition for the symplectic eigenvalue $z_2$ :
\begin{equation}
z_2\geq \Big|\frac 1 2-x\Big|.
\end{equation}
Now, the sum of symplectic eigenvalues gives:
\begin{equation}
  z_1+z_2\geq |x|+\Big|\frac 1 2 -x\Big|\geq \Big|x+\frac 1 2 -x \Big|=\frac 1 2.
\end{equation}
The above inequality assures that the condition (\ref{wine}) is always fulfilled. The generalization to more modes is straightforward. For instance, consider a three-mode CM and we want an EW detecting three-partite entanglement. Then, according to the Proposition, we need to impose constraints on the three block diagonal matrices of the witness, which amount to the following inequalities for the corresponding symplectic eigenvalues:
\begin{eqnarray}\label{gen}
& z_1\geq |x_1|, \quad x_1\in \mathbb{R},\nonumber\\
& z_2\geq |x_2|, \quad x_2\in \mathbb{R},\\
&  z_3\geq  \Big|\frac 1 2 -x_1-x_2\nonumber\Big|.
\end{eqnarray}
These inequalities imply the constraint (\ref{wine}).
\item In the second part, we present the generalization of the proof for $k-$partite entanglement of $N-$mode CMs, with $k<N$. Consider, for simplicity, a three-mode state and the bipartition between the first and the other two modes. The witness $Z$ is a $6\times 6$ matrix where $Z_1$ is the $2\times 2$ block diagonal matrix of $Z$ acting on the first mode, and we denote by $Z'$ the $4\times 4$ block matrix acting on the other two modes. Then the corresponding constraints on the witness are:
\begin{eqnarray}
& Z\geq 0, \nonumber\\
& Z_1+i x \Omega_{1} \geq 0, \quad x\in \mathbb{R},\\
& Z'+i \frac{1}{2}(\frac 1 2-x) \Omega_{2} \geq 0.\nonumber
\end{eqnarray}
If we denote by $z_1$ the symplectic eigenvalue of $Z_1$, and by $z_1'$, $z_2'$ the two symplectic eigenvalues of $Z'$, then the conditions above are equivalent to:
\begin{eqnarray}
& z_1\geq |x|, \quad x\in \mathbb{R},\nonumber\\
& z_1'\geq \frac 1 2 \Big|\frac 1 2 -x\Big|, \\
&  z_2'\geq \frac 1 2 \Big|\frac 1 2 -x\Big|\nonumber,
\end{eqnarray}
which satisfy the condition (\ref{wine}). Since this is a bipartite state, the lower bounds for the three symplectic eigenvalues depend on a single parameter $x$, while, according to the Proposition, the detection of three-partite entanglement in a three-mode state would require two optimization parameters, $x_1$ and $x_2$ (see Eq. (\ref{gen})). The generalization of the proof to $N$ modes and $k$ parties is straightforward. Note that the conditions in the Proposition are stronger for $k-$partite entanglement (with $k<N$) than for genuine multipartite entanglement (i.e. $k=N$), where the optimization for every symplectic eigenvalue is done independently.
\end{enumerate}\end{proof}

While the semidefinite inequalities proposed in the previous Proposition present the advantage of being linear, the drawback of these constraints is that they are stronger than those required by the Theorem characterizing the EWs based on second moments, and therefore some EWs will not satisfy conditions (\ref{prop}).

\section{Entanglement witnesses from random measurements}

Here we will shortly present the physical set-up of the homodyne detection, that encodes the experimental settings measuring the variances of the state. Homodyne measurements are phase sensitive measurements which allow the detection of the moments of quadratures up to the second order \cite{braun,weed}. We denote by $\hat k$ and $\hat k^\dag$ the mode operators of our state. A simple scheme for balanced homodyne measurements is composed of a balanced beam splitter superposing the signal mode to be measured $\hat k$ with a strong local oscillator field $\alpha_{LO}=|\alpha_{LO}|e^{i\theta}$ with phase $\theta$, and two photon detectors, converting the electromagnetic modes into two output photon currents, $i_1$ and $i_2$. The actual quantity to be measured is the difference in the photon currents, given by:
\begin{equation}
\delta i =i_1-i_2=q |\alpha_{LO}| \ \langle\hat x_{\theta}\rangle,
\end{equation}
with $q$ being a constant, and $\hat x_{\theta}$ is the generalized quadrature operator of mode $\hat k$ defined as:
\begin{equation}
  \hat x_{\theta}=\frac{\exp{(-i\theta)}\hat k+\exp{(i\theta)}\hat k^{\dag}}{\sqrt{2}},
\end{equation}
which covers the whole continuum of quadratures for $\theta \in [0, \pi]$. It was shown in Ref. \cite{proj} that in the strong local oscillator limit the homodyne detection performs the projective measurements corresponding to POVM $|x_\theta\rangle \langle x_\theta|$, where $| x_\theta\rangle$ is the eigenstate of the quadrature phase operator $\hat x_\theta$.

In two-mode homodyne detection, we rely on the detection scheme proposed in Ref. \cite{dauria}, where the two-mode states are characterized by a single homodyne detector. By denoting with $\hat a$ and $\hat b$ the initial modes to be detected,  the mode $\hat k$ arriving at the detector can be expressed as \cite{dauria}
\begin{equation}\label{kk}
  \hat k=\exp(i \varphi) \cos \phi \ \hat a +\sin \phi \ \hat b,
\end{equation}
which corresponds to applying a phase shift of angle $\varphi$ between the horizontal and vertical polarization components, a polarization rotator of angle $\phi$, and a polarizing beam splitter (PBS) reflecting the vertically polarized component of the beam toward the detector \cite{dauria}. Using repeated measurements of the quadratures for a set of identical states, the homodyne data are collected for which a probability distribution can be assigned with the variance given by:
\begin{equation}\label{p}
\langle \hat x_{\theta}^2 \rangle-\langle \hat x_{\theta}\rangle^2=\Tr[P \gamma],
\end{equation}
where $P$ is the matrix for the measurement of the quadrature variance of the mode $\hat k$:
\begin{equation}
P=u u^{\rm T}, u^{\rm T}=\left({\begin{array}{*{20}c}
\cos \phi \cos(\theta-\varphi)& \cos \phi \sin(\theta-\varphi)&\sin \phi \cos \theta & \sin \phi \sin \theta \end{array}}\right).
\end{equation}
As $P$ is a symmetric, real $4\times4$ matrix we can see that for $10$ different combinations of angles $\theta$, $\phi$ and $\varphi$ the entire two-mode CM can be reconstructed (the number of unknown independent parameters in an $N-$mode CM is $N(2N+1)$).

The extension of detection to $N-$mode CV states by a single homodyne detector can be achieved by applying the same two-mode combination scheme $N-1$ times. For example, for the initial modes $\hat a$, $\hat b$ and $\hat c$, the generalized mode arriving at the detector is:
\begin{equation}
\hat k= \exp(i \varphi_1) \cos \phi \  \hat a + \exp(i \varphi_2)\sin \phi \cos \psi \ \hat b+ \sin \phi \sin \psi \  \hat c,
\end{equation}
from where we can see that for $\psi=0$ and $\varphi_2=0$ the two-mode case in Eq. (\ref{kk}) is obtained.
We denote by $P_j$ the matrix of the j-th measurement.

\subsection{Constructing witnesses}
Random measurement directions in the case of two modes are given by random angles $\theta, \phi, \varphi$ that are drawn from a uniform distribution in an interval:
\begin{eqnarray}\label{ang}
  0\leq & \theta\leq \pi, \\
  0\leq & \phi \leq \pi, \\
  0\leq & \varphi < 2 \pi.
\end{eqnarray}

The problem of finding a witness operator $Z$, given the repeated independent measurements $P_j$ on the CM, reduces to finding the coefficients $c_j$ such that $Z=\sum_{j} c_j P_j$. Therefore, we apply the Proposition in order to find the best witness for two-mode CMs, and propose the following SDP:

\begin{eqnarray}
{\rm minimize~over} {~x:} {~~~~}{\bf{c\cdot m}}{}{}\nonumber \\
{\rm subject~to:}~~ {Z=\sum_j c_j P_j,}\nonumber \\
~~~~~~~~~~~~~~~~~{Z=\left({\begin{array}{*{20}c}
      Z_1 & Z_c \\
      Z_c^{\rm T} & Z_2 \end{array}}\right)\geq 0, }\\
~~~~~~~~~~~~~~~~~{Z_1+i x \Omega_1 \geq 0, }\nonumber \\
~~~~~~~~~~~~~~~~~{Z_2+i (\frac 1 2 - x) \Omega_1 \geq 0}, \nonumber~\label{min}
\end{eqnarray}
where $\bf{m}=\Tr(P\gamma)$, with $\bf{P}$ being the vector of measurement matrices $P_j$. This SDP finds the matrix $Z$, given the experimentally obtained data, such that
\begin{equation}\label{w}
  \bf{c\cdot m}=\Tr[Z\gamma]
\end{equation}
takes its minimal value, while being an EW as defined in Theorem above. If the obtained value in Eq. (\ref{w}) is smaller than one, then the CV state with CM $\gamma$ can be unambiguously identified as being entangled.

This SDP also allows for the identification of the minimal number of measurements that are required for entanglement assessment in arbitrary states. The number of measurements in a tomographically complete setting is given by $N(2N+1)$, where $N$ is the number of modes. This is the maximal number of measurement settings required to detect entanglement. However, the set of EWs described in the Proposition is more restrictive than the set of all EWs. The consequences will be discussed later.

\section{Detection of non-PPT entanglement and bound entanglement}

The proposed SDP has the immediate advantage that it does not require any information about the state, except the number of modes $N$. We will now test the performance of this method by simulating its implementation on random two-mode entangled CV states, and on four-mode bipartite bound entangled states. \\
\indent The entanglement of two-mode CMs with block structure given by:
\begin{equation}\label{bl}
\gamma=\left({\begin{array}{*{20}c}
\gamma_1 & \varepsilon_{1,2}\\
\varepsilon_{1,2}^{\rm T} & \gamma_2
\end{array}}\right),
\end{equation}
is quantified by means of the logarithmic negativity \cite{ser4}:
\begin{eqnarray}\label{ent}
E=\max \{0,-\frac{1}{2}\log_2 f\}, \end{eqnarray} where \begin{eqnarray}\label{logneg}
f=\frac{1}{2}(\det \gamma_1 +\det
\gamma_2)-\det \varepsilon_{1,2}
-\left({\left[\frac{1}{2}(\det \gamma_1+\det \gamma_2)-\det
\varepsilon_{1,2}\right]^2-\det\gamma}\right)^{1/2}.\nonumber\end{eqnarray}
An EW provides a lower bound for the logarithmic negativity measure when the positive partial transpose (PPT) criterion of separability is necessary and sufficient \cite{and}:
\begin{equation}
E\geq \log_2 \frac{1}{w},
\end{equation}
where $w\in(0,1)$ is the outcome of measuring an EW on CM $\gamma$: $\Tr[Z\gamma]=w$. For two-mode CMs the logarithmic negativity corresponds to the minimal\footnote{Compared to the optimal EW in state space, the minimal EW based on second moments gives the best estimate of the degree of entanglement the
considered state has, but it is not necessarily the finest witness \cite{and}. } EW $Z_{min}$ giving the smallest possible value $w_{min}$.

In the following we investigate the efficiency of our method for detecting entanglement of arbitrary CV states, with respect to the minimal number of measurements required to accomplish this task. Thus, given an arbitrary unknown CM our algorithm first computes the variances of the generalized quadrature (\ref{p}) for one random measurement direction in phase space and then carries out the SDP optimization to check if the state is entangled. If entanglement is not detected,  additional random measurements are successively simulated and the optimization algorithm is executed each time until the entanglement is detected. At least two measurement settings are required in order to detect entanglement.

\subsection{Detecting entanglement in squeezed vacuum states}

The CMs of squeezed vacuum states have the form:
\begin{equation}\label{sts}
\gamma=\left({\begin{array}{*{20}c}
\cosh 2r & 0 &  \sinh 2r & 0 \cr
0 & \cosh 2r & 0 & - \sinh 2r \cr
 \sinh 2r & 0 & \cosh 2r & 0 \cr
0 & - \sinh 2r & 0 & \cosh 2r
\end{array}}\right),
\end{equation}
where $r$ is the squeezing parameter. For such states the logarithmic negativity can be calculated using (\ref{ent}), obtaining a linear dependence on the squeezing parameter:
\begin{equation}\label{entv}
 E=2 ~  r \log_2 e,\end{equation}
where $e$ is the Euler constant. Squeezed vacuum states are Gaussian states, which are naturally accessible in many experimental situations where spontaneous down conversion is involved, being also useful in many quantum optics applications \cite{braun, weed} \footnote{Recent experiments report the achievement in measuring 15 dB of squeezed light \cite{vahl}, which corresponds to $r\approx 1.73$ according to the formula \cite{ads}: $ \# {\rm dB}= 10 \log_{10} e^{2 r}$.}.
\begin{figure}[h]
\centering
\includegraphics[width=0.95\columnwidth]{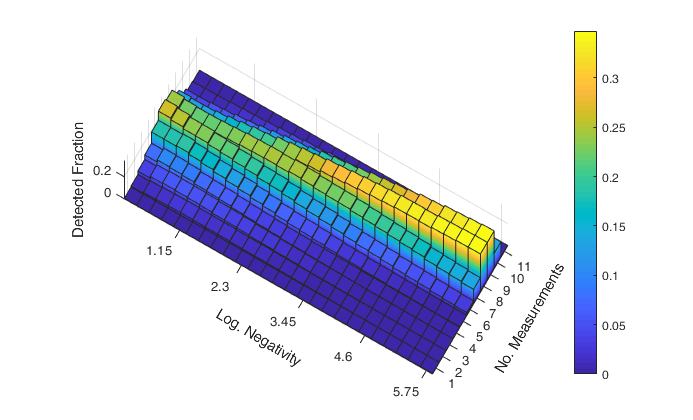}\\
\caption{Fraction of entanglement detection of squeezed vacuum states: $5\times10^5$ runs of the algorithm on the two-mode squeezed vacuum states (\ref{sts}) with squeezing parameter $r\in[0,2]$. The logarithmic negativity is given by $E=2 ~  r \log_2 e$ (see Eq. (\ref{entv})). By successively adding measurement directions, the EW is evaluated at every round until the presence of entanglement is certified. The data are normalized such that they sum up to $1$ for every value of entanglement.}
\end{figure}

In Fig. 1 we show the fraction of entanglement detection of squeezed vacuum states, using random EWs. Contrary to intuition, states with less entanglement are more easily detected, i.e. they require on average fewer measurements than states with higher entanglement. This is due to the fact that in this case the amount of entanglement is linked to the strength of  quadrature squeezing. It is well known that it is difficult to measure high squeezing in CV states \cite{vahl} (see also the explanation given in Fig. 2). The full tomography for two-mode CMs is reached by $10$ independent measurements. The CM  (\ref{sts}) of the squeezed vacuum state has some zero elements, and with this knowledge about the state one would need only $6$ measurements to reconstruct the CM entirely. However, our method may require more than $6$ measurements to assess entanglement since we assumed no information about the states, except the dimension of the CM. As a consequence of the stronger constraints imposed on the EWs in Eqs. (\ref{prop}) our method requires, with very low probability  ($0.0094 \%$ in our example), more than $10$ measurements, which correspond to full tomography.\\
\begin{figure}[h]
\centering
\includegraphics[width=0.48\textwidth]{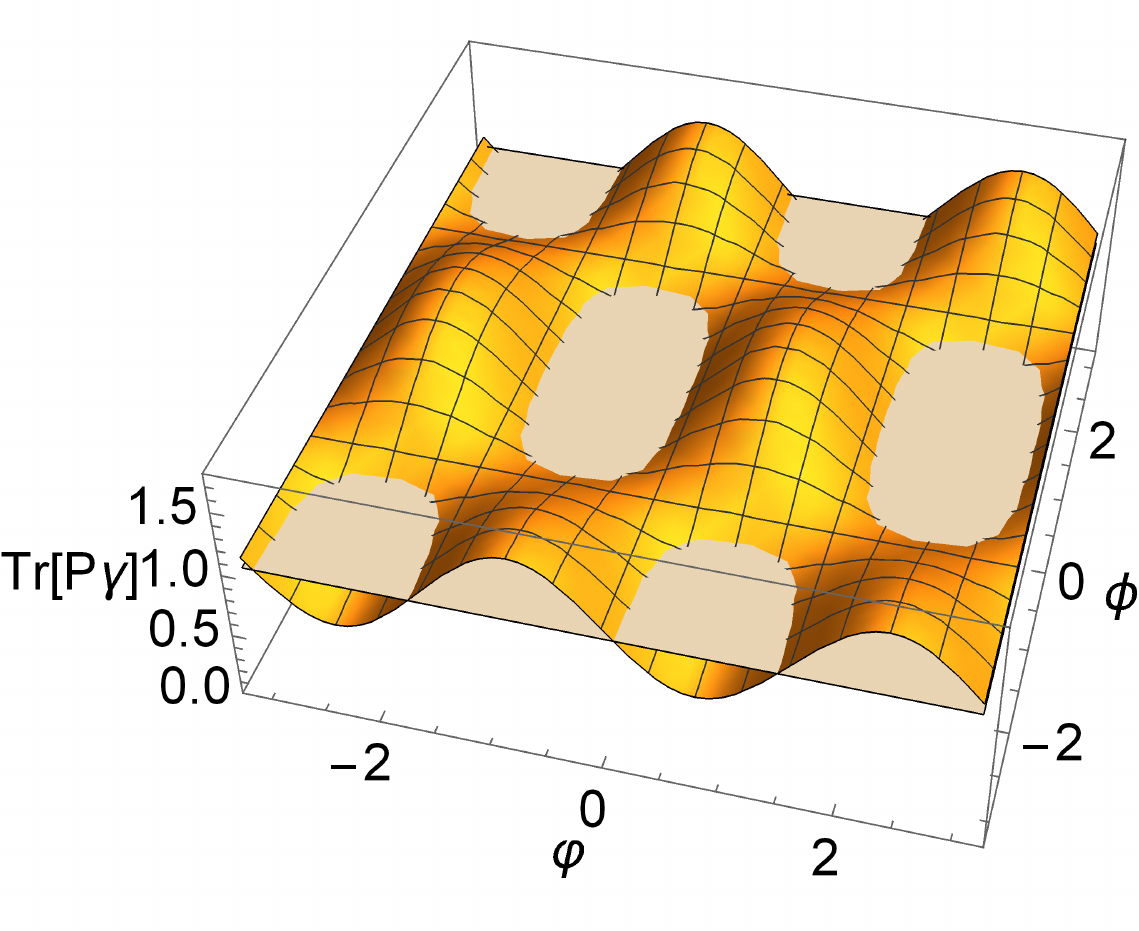}\qquad \quad
\includegraphics[width=0.48\textwidth]{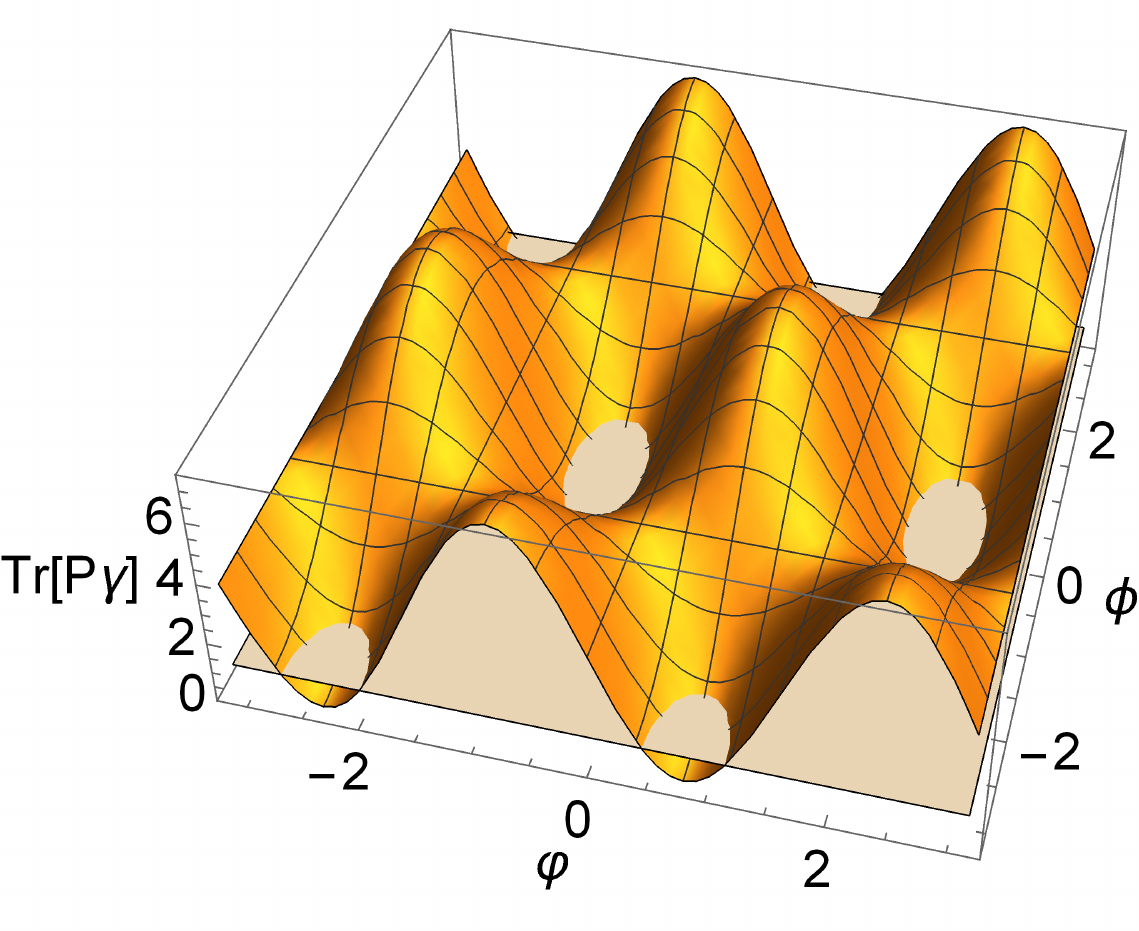}
\caption{ The variance of the generalized quadrature $\Tr[P \gamma]$, see Eq. (\ref{p}), of the squeezed vacuum CM, as a function of $\varphi\in [-\pi,\pi]$ and $\phi\in [-\pi,\pi]$, for $\theta=0$ and squeezing parameter $r=0.2$ (left) and $r=1$ (right). The horizontal plane represents $\Tr[P \gamma]=1$, which is the case of the squeezed vacuum states with $r=0$ (separable states). In the regions below this plane entanglement is detected.}
\end{figure}

In Fig. 2 we show the variance of the generalized quadrature $\Tr[P\gamma]$, see $(\ref{p})$, for $\theta=0$, as a function of $\varphi\in [-\pi,\pi]$ and $\phi\in [-\pi,\pi]$, for different values of the squeezing parameter, $r=0.2$ (left) and $r=1$ (right), of the squeezed vacuum state. The outcomes of the random measurements are represented by the points on this surface. The horizontal plane is given by $\Tr[P \gamma]=1$, which holds for a separable vacuum state with $r=0$. The areas below this plane, where $\Tr[P \gamma]<1$, correspond to the region of parameters $\varphi$ and $\phi$ for which entanglement is detected. We observe that the areas of the regions of entanglement detection are decreasing with increasing the squeezing. This corresponds to the fact that highly squeezed states occupy a smaller region in phase space in terms of the angles $\phi, \varphi$. Thus, more random measurements are needed to detect the entanglement.

\subsection{Detecting entanglement in random covariance matrices}

Random CMs are produced as follows. Starting with a CM in diagonal form, with symplectic eigenvalues $\nu_i\geq1$ ($i=1,...,N$) randomly generated from a uniform distribution in a finite real interval $[1,t]$, $t>1$:
\begin{equation}\label{th}
\gamma_{th}= \bigoplus_{i=1}^N \left({\begin{array}{*{20}c} \nu_i & 0 \\ 0 & \nu_i \\ \end{array}}\right),
\end{equation}
the general random CMs $\gamma$ are created by applying random symplectic transformations $S\in Sp(2N,\mathbb R)$, as follows \cite{sera}:
\begin{equation}\label{th1}
\gamma=S  \gamma_{th} S^{\rm T}.
\end{equation}
The matrix (\ref{th}) is the CM of thermal states, with the symplectic eigenvalue of every mode $i$ related to the thermal photon number $n_i$ as follows: $\nu_i=2 n_i +1$ \cite{sera}. Random symplectic matrices are generated using the Euler decomposition (\ref{eu}). First, unitary matrices $X$ and $Y$ in Eq. (\ref{ku}) are generated from the Haar distribution \cite{dutta}, and the symplectic orthogonal matrices $K$ and $L$ are formed as in Eq. (\ref{sorth}). The one-mode squeezers defined in Eq. (\ref{sqr}) are created by randomly choosing parameters $r_i$ via a uniform distribution in a finite interval. For this purpose we implemented the Matlab code presented in Ref. \cite{jagger}.

In Fig. 3 we illustrate the efficiency of entanglement detection for general random two-mode CMs, created from thermal state CMs (\ref{th}) with random symplectic eigenvalues $\nu_i\in[0,5]$, by random symplectic transformations (\ref{eu}) with squeezing parameters $r_i\in[0,2]$. The probability that entanglement is detected by $11-12$ measurements in this case, is $0.05 \%$. Our method shows a slight improvement in the efficiency of entanglement detection for highly entangled states compared to less entangled states, and most of the time it does not require full tomography.

The evident difference in the efficiency of entanglement  detection in random CMs compared to squeezed vacuum states may reside in the fact that highly squeezed states look classical in random measurement directions, which does not have to be the case for random states. In addition, squeezed vacuum states are a special class of states for which the logarithmic negativity has a linear dependence on the squeezing parameter alone (see Eq. (\ref{entv})), while for a general two-mode CM the logarithmic negativity depends also on thermal photon number of the modes, and the simulation of entanglement detection shows a different behaviour.

In general, it is unlikely to draw randomly an entangled state with high logarithmic negativity, especially for states with a high number of modes. However, for the two-mode CMs, with the range of entanglement considered in Fig. 3, a substantial fraction of randomly generated CMs is entangled, which allowed us to perform the simulation.
\begin{figure}[h]
\centering
\includegraphics[width=0.95\columnwidth]{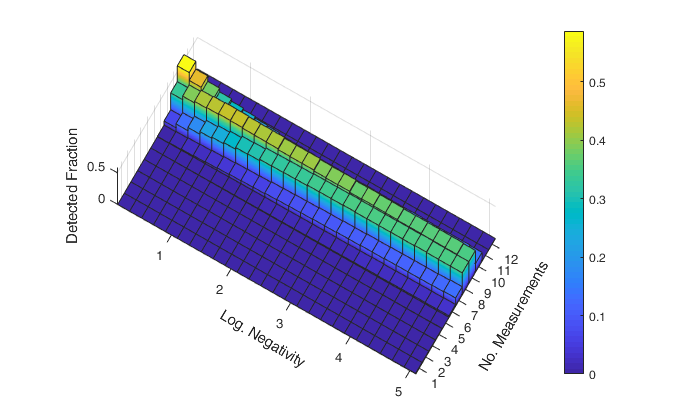}\\
\caption{Fraction of entanglement detection for random two-mode states: $5\times10^5$ runs of the algorithm on random two-mode CMs for $\nu_i\in[0,5]$ and $r_i\in[0,2]$. By successively adding measurement directions, the EW is evaluated at every round until the presence of entanglement is certified. The data are normalized such that they sum up to $1$ for every value of entanglement.}
\end{figure}

\subsection{Detecting bipartite bound entanglement}

Since the proposed SDP algorithm can be easily generalized to multi-mode CV states, we provide an example of a four-mode CM with $12$ independent parameters, mentioned in Ref. \cite{wolf}, which has bipartite bound entanglement:
\begin{equation}\label{4}
\gamma=\left({\begin{array}{*{20}c}
2 & 0 & 0 & 0 & 1 & 0 & 0 & 0\cr
0 & 1 & 0 & 0 & 0 & 0 & 0 & -1\cr
0 & 0 & 2 & 0 & 0 & 0 & -1 & 0\cr
0 & 0 & 0 & 1 & 0 & -1 & 0 & 0\cr
1 & 0 & 0 & 0 & 2 & 0 & 0 & 0\cr
0 & 0 & 0 & -1 & 0 & 4 & 0 & 0\cr
0 & 0 & -1 & 0 & 0 & 0 & 2 & 0\cr
0 & -1 & 0 & 0 & 0 & 0 & 0 & 4
\end{array}}\right).
\end{equation}
The detection of bound entanglement by our method proves that the EWs defined in the Theorem above, goes beyond the criteria which detect entanglement only in states with non-positive partial transpose. A general $N$-mode CM has $N(2N+1)$ independent variables, and for the four-mode CM in Eq. (\ref{4}) by performing $36$ measurements our algorithm provides the best estimate of entanglement, $\Tr[Z_{min} \gamma]=0.8966$, which is in agreement with the results of Ref. \cite{hyll}. In Fig. 4 we depict the frequency of entanglement detection as a function of the number of random measurements composing the witness. The CM in Eq. (\ref{4}) is of a rather simple form, however, the construction of the EW detecting bound entanglement requires $33$ random measurements on average, since our SDP considers the number of modes of the state as the only available information.
\begin{figure}[h]
\centering
\includegraphics[width=0.8\columnwidth]{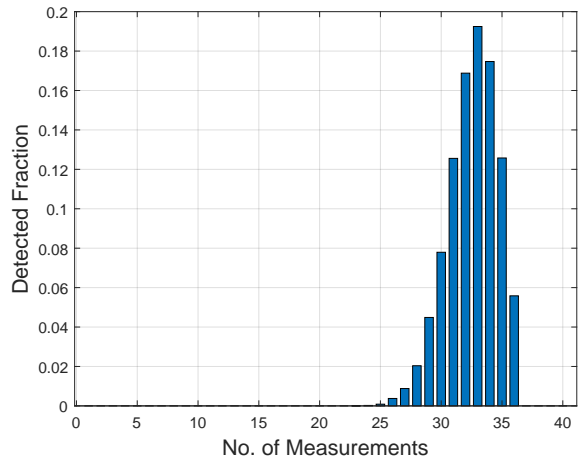}\\
\caption{Fraction of entanglement detection for $4-$mode bipartite bound entangled state, see  Eq. (\ref{4}): $10^4$ runs of the algorithm. The data are normalized such that they sum up to $1$.}
\end{figure}

\section{Statistical analysis}

Until now we have considered only ideal measurements, where we used the exact variances $m_i={\rm Tr}[P_i \gamma]=(\Delta \hat{x} _{\theta_i})^2$  (see Sec. 4) in order to construct the entanglement witness. In real experiments the accessible data are subject to statistical fluctuations. In the following we perform the statistical analysis for the case of Gaussian states, that is,  we assume that the data  obtained in homodyning, which represent the collection of outcomes $X_{ij}=\langle \hat{x}_{\theta_i}\rangle_j$, ($j=1,\ldots,n_i$), from $n_i$ repetitions of the measurement with the measurement direction given by $\theta_i$, are governed by the normal probability distribution $\mathcal{N}_i(\mu_i,m_i)$ with the mean $\mu_i$, and variance $m_i=(\Delta \hat x_{\theta_i})^2$. Given the homodyne data from $n_i$ measurements for a fixed measurement direction $\theta_i$, the sample variance denoted as $\bar P_i$, which estimates the variance $m_i$, is given by \footnote{Using $n_i-1$ instead of $n_i$ corrects the bias in the estimation of the population variance, and is called Bessel's correction \cite{ken}. This method is necessary when the population mean $\mu_i$ is unknown, but is estimated by the sample mean $\bar X_i$. }:
\begin{equation}\label{sampv}
\bar P_i=\frac{1}{n_i-1} \sum_{j=1}^{n_i} (X_{ij}-\bar X_{i})^2,
\end{equation}
where $\bar X_{i}$ is the sample mean:
\begin{equation}
 \bar X_{i}=\frac{1}{n} \sum_{j=1}^{n_i} X_{ij}.
\end{equation}
In this case, the estimated value of our witness ${\rm Tr}[Z \gamma]$, denoted as $\bar Z$, is given by:
\begin{equation}\label{zp}
 \bar Z=\sum_i c_i \bar P_i,
\end{equation}
where the index $i$ is used to denote different measurement settings, and the coefficients $c_i$ were introduced in Eq.(\ref{min}). In the case when the data comes from a Gaussian probability distribution, the distribution of the sample variance follows the $\chi_{n_i-1}^2$ distribution \cite{chi}:
\begin{equation}
 \frac {n_i-1} {m_i}\bar P_i\sim  \chi_{n_i-1}^2,
\end{equation}
where $\chi_{n_i-1}^2$ is the chi-square distribution with $n_i-1$ degrees of freedom, which by definition represents the distribution of sum of squares of $n_i-1$ independent, standard normal random variables. The statistical error carried by $\chi_{n_i-1}^2$ is given by:
\begin{equation}
\Delta \chi_{n_i-1}^2=\sqrt{{\rm Var}(\chi_{n_i-1}^2)}=\sqrt{2(n_i-1)},
\end{equation}
where ${\rm Var}(\chi_{n_i-1}^2)=2(n_i-1)$ is the variance of the chi-square distribution \cite{chi}. Using the error propagation formula the uncertainty of $\bar P_i$ satisfies:
\begin{equation}\label{inneq}
 \Delta \bar P_i=\frac{d \bar P_i}{d \chi_{n_i-1}^2} \Delta \chi_{n_i-1}^2=\frac{m_i}{n_i-1}\Delta \chi_{n_i-1}^2=m_i\sqrt{ \frac{2}{n_i-1} }.
\end{equation}
Using again standard error propagation and considering that the number of measurement repetitions is equal for every measurement direction, i.e., $n=n_i$ for every $i$, we obtain that the resulting error of $\bar Z$ defined in Eq. (\ref{zp})  has the following expression:
\begin{equation}\label{indep}
  \Delta \bar Z=\sqrt{\sum_i \Big(\frac{d \bar Z}{d \bar P_i}\Big)^2 (\Delta \bar P_i)^2}=\sqrt{ \frac{2}{n-1} }\sqrt{\sum_i c_i^2 m_i^2}.
\end{equation}
We stress the fact that, although by our method we can also detect entanglement in non-Gaussian states, this formula for the error of the value of the witness is valid only for Gaussian states. If the $X_{ij}$ are not normally distributed, then the statistical analysis of entanglement witnesses based on second moments will require also higher moments of the distribution.

In our method the coefficients $c_i$ are derived from the variances $m_i$ (see Eq. (\ref{min})), while Eq. (\ref{indep}) neglects the fact that they are not independent. To solve this difficulty one has to divide the homodyne data into two sets. First, one of them is used to derive the coefficients $c_i$, and then this witness is evaluated using the variances obtained from the other set of data \cite{mor}. In this way, the coefficients can be regarded as independent from the errors in the variances of the second set of data.
\begin{figure}[h]
\centering
\includegraphics[width=0.7\columnwidth]{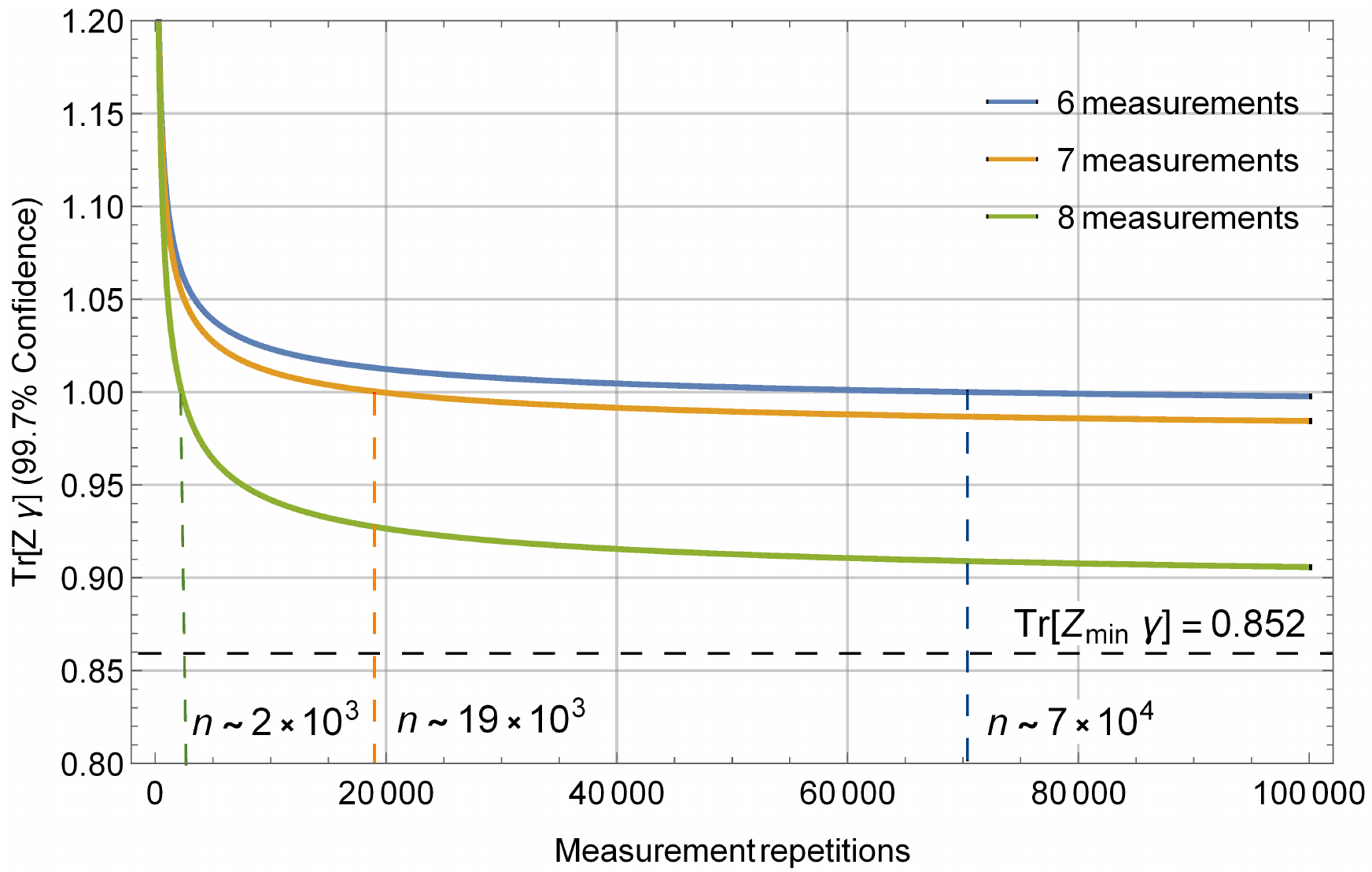}\\
\caption{The maximum of the $3\sigma$ confidence interval for the witness $Z$ of a Gaussian CM $\gamma$, as obtained by the statistical estimate according to Eq. (\ref{indep}). The horizontal dashed line indicates the minimal value of the witness for the considered CM $\Tr[Z_{min}\gamma]=0.852$. The vertical dashed lines indicate the number of measurement repetitions required to detect entanglement with $6$ (blue), $7$ (orange) and $8$ (green) measurement settings.}
\end{figure}
With the quantity in Eq. (\ref{indep}) it is possible to decide whether it is better to perform additional repetitions of the measurements, or to add new measurements to detect entanglement. For example, consider the single detection of a low entanglement CM with $\Tr[Z_{min} \gamma]=0.852$. In Fig. 5 the $3\sigma$-confidence of $\Tr[Z\gamma]$ is plotted as a function of the number $n$ of measurement repetitions. It shows that a certification of $\Tr[Z\gamma]<1$ with $99.7\%$-confidence is possible for $6$ measurements, which requires a high number of repetitions of the measurements. However, this number significantly decreases when adding another measurement setting.

\section{Summary and conclusions}

We have proposed a method to detect entanglement of unknown CV states, given only the dimension of their covariance matrices, using random homodyne measurements. Our method provides an alternative for performing full tomography. We characterize the entanglement witnesses based on second moments using stronger semidefinite constraints than those presented in Ref. \cite{hyll}, and which account for obtaining a valid witness at all times. Therefore, a quantum state can be clearly considered entangled if it is detected by this criterion. As these constraints are linear, they can be implemented in an SDP. We studied the feasibility of this method in experimental situations, where the figure of merit is considered the number of measurements required to detect entanglement.

First, we tested the proposed algorithm for two-mode squeezed vacuum states, for which the logarithmic negativity linearly depends on the squeezing. We showed that the number of necessary random measurements is very likely to be smaller than for full tomography. We observed an increasing number of measurements required to detect highly entangled states, which is explained by the well-known difficulties in detecting high squeezing.

Our primary objective was to simulate the performance of this method for uniformly drawn random  two-mode covariance matrices. Without adding any information about the states, we still found a reduction in the number of measurements needed to certify the presence of entanglement. The phenomenology of entanglement detection in random CV states is very similar to the case of decomposable witnesses for discrete systems \cite{jochn}. Hence, a higher entanglement is easier to detect, but in our case this improvement is not as significant as in the discrete case. Only with low probability our method needs a tomographically complete set of measurements to detect entanglement in random two-mode states.

Bound entangled CV states can also efficiently be detected by a random entanglement witness. Similarly to the previous cases entanglement is detected with less than a tomographically complete set of measurements.

The experimental scheme implementing our method for two-mode CV states consists of a phase shift in polarization basis, a rotator of polarization, a polarizing beam splitter and a homodyne detector, as e.g. presented in Ref. \cite{dauria}. We also extended this scheme to multimode CV states. We investigated the statistical robustness of the method, and showed that it has a good robustness to statistical errors.

\paragraph{Acknowledgements.} The authors thank Jochen Szangolies, Thomas Wagner and Matteo Paris for helpful discussions.
Giulio Gianfelici has received funding from the German Federal Ministry
of Education and Research (BMBF), within the Hardware-based Quantum
Security (HQS) project.

\end{document}